\documentstyle[aps,preprint]{revtex}

\begin{document}
\draft
\title{
 Non-magnetic impurities in $S=1/2$ spin-Peierls system and Heisenberg
ladder systems
}
\author{Tai-Kai Ng}
\address{
Department of Physics,
Hong Kong University of Science and Technology,\\
Clear Water Bay Road,
Kowloon, Hong Kong
}
\date{ \today }
\maketitle
\begin{abstract}
 In this paper we study the effect of non-magnetic impurities in
spin-Peierls system $CuGeO_3$ within the framework of non-linear
sigma model plus topological term. We show that local moments
are induced in both the $Zn$-doped and $Si$-doped $CuGeO_3$
compounds. Effective low energy theories for the impurity-induced
local moments are derived in both cases where interesting 
differences between the two cases and between the $Zn$-doped
Heisenberg two-ladder system are pointed out. The low-energy
physics of the $Zn$-doped Heisenberg three-leg ladder system
is also discussed. 

\end{abstract}


\narrowtext

 Recently, there are lot of interests in the study of 
quasi- one dimensional spin $1/2$ Heisenberg spin systems,
including the spin-Peierls system $CuGeO_3$\cite{edimer} and 
Heisenberg ladder systems $Sr_nCu_{n+1}O_{2n+1}$\cite{el2,rice}.
In particular, a lot of efforts have been spent on the study of
impurity effects on $CuGeO_3$ and $SrCu_2O_3$ compounds where
surprising results were observed. In $CuGeO_3$, which is an
inorganic spin-Peierls system, it was observed that local magnetic
moments were induced when a few precent of $Cu$ is replaced by 
$Zn$\cite{ed1}, or when $Ge$ is replaced by $Si$\cite{ed2}. In
the case of $Cu_{1-x}Zn_xGeO_3$, it was discovered that the
spin-Peierls state collapses at around $x=0.03$, and is replaced 
by a new magnetic state which was interpreted as a 
spin-glass-like state\cite{ed1} or long-range antiferromagnetic
state\cite{oser}. In $CuGe_{1-y}Si_yO_3$, long-range antiferromagnetic
order co-existing with spin-Peierls state was also observed 
with very small amount of disorder, when $y=0.007$\cite{neu,pha}.
Similar 'disorder induced' local magnetic moments were also
observed in the two-leg ladder compound $Sr(Cu_{1-x}Zn_x)_2O_3$
when $Cu$ ions are replaced by $Zn$ ions\cite{az1} and 
long-range antiferromagnetic order was observed at
low temperature at $x\sim0.07$\cite{az2}. The possibility of
co-existing spin-Peierls and long-range antiferromagnetic state
and formation of local moments in $Zn$-doped two-leg ladder system
has been investigated by bosonization technique\cite{fu,na}. It
was also suggested that the low-energy physics of the $Zn$-doped
Heisenberg ladder system can be described by a $S=1/2$ random J 
Heisenberg spin chain where the nearest-neighbor coupling J has 
both random sign and magnitude\cite{si}. 

   In this paper we shall discuss the effect of non-magnetic impurities
in the spin-Peierls system within the framework of
non-linear sigma model ($NL\sigma{M}$) plus topological terms
\begin{equation}
\label{nlsm}
{\em L} = {1\over{g}}(\partial_{\mu}\vec{n})^2+i\lambda(|\vec{n}|^2
-1)-i{\theta\over8\pi}\epsilon^{\mu\nu}\vec{n}.(\partial_{\mu}\vec{n}
\times\partial_{\nu}\vec{n}),
\end{equation}
where $\vec{n}$ is a unit vector and $g\sim2/S=4$. The model
is believed to be the correct starting point for Heisenberg spin 
chains\cite{hal,aff}. We shall show that the formation of
magnetic moments in both $Zn$-doped and $Si$-doped $CuGeO_3$ 
compounds can be explained within 
the $NL\sigma{M}$ treatment. The corresponding low energy
theories for finite concentration of impurities will be
derived where similarities and differences between the two cases
and between the $Zn$-doped Heisenberg
two-ladder system will be pointed out. The case of $Zn$-doped
Heisenberg three-leg ladder system will also be discussed.
We shall argue that the experimentally
observed 'disorder induced' magnetic states in different
compounds are consistent with our effective theories.

   In the $NL\sigma{M}$ formulation, a dimerzied $S=1/2$ spin chain with
alternating interaction $J_{i,i+1}=J[1+\gamma(-1)^i]$ can be described
for small $\gamma$ by a $NL\sigma{M}$ with with topological term
$\theta=\pi(1+\gamma)$\cite{aff}. For an infinite chain, there is
a spin gap $\Delta_g\sim{J}\gamma^{2/3}$ in the excitation spectrum
\cite{cf}. We shall assume that the only effect of substituting $Cu$  
by $Zn$ is to remove spins randomly from the spin chain,
leading to a collection of open spin chains with average length
$\sim{x}^{-1}$, where $x$ is the $Zn$ concentration.

   The properties of open spin chains have been studied by the
author\cite{ng94} where formation of boundary excitations (end states)
and their topological characters were discussed.  For open $S=1/2$
dimerized spin chain with even number of sites $i=1,2m$, it was 
found that localized $S=1/2$ excitations exist at the ends of
spin chain when $\gamma>0$, but are absent otherwise. The
existence of end states when $\gamma>0$ can be understood most easily
in the limit $|\gamma|\rightarrow1$. In this limit, the spin chain
is completely dimerized and no end states appear if $J_{i,i+1}$ is
finite for {\em odd} $i$ and zero for {\em even} $i$ ($\gamma<0$). 
In the other case when $J_{i,i+1}$ is nonzero when $i$ is {\em even}
($\gamma>0$), the first and last sites are decoupled from the rest 
of the spin chain forming the $S=1/2$ end states. For $|\gamma|$
small, the end spins are not completely decoupled from the spin
chain and end states which decay into the spin chain with finite
correlation length $\xi\sim\Delta_g^{-1}\sim|\gamma|^{-3/2}$ 
are formed\cite{ng94}. For the randomly doped $Cu_{1-x}Zn_xGeO_3$ 
compound, it is expected that open spin chains with both even
and odd number of sites are formed. However, the polarization of 
the dimerization $\gamma$ is not determined a priori but is
determined by the underlying dynamics of the broken spin chain.
In particular, we expect that even chains will always
dimerized with $\gamma<0$ to maximize the number of short-range
spin-singlet bonds and no end states are formed. However, the
situation is different for spin chains with odd number of
sites where the ground state is expected to be a spin doublet
with total spin $S_{tot}=1/2$. To respect parity, we
expect that the spin chain will be dimerized in a symmetric
way leaving an 'extra' spin $S=1/2$ as a localized state
at the middle of spin chain. Notice that in this case, the
spin chain has opposite polarization of dimerization at 
two ends of the spin chain, and the localized $S=1/2$ state at
the middle of the spin chain is just a soliton solution
at the domain wall seperating the two regions of polarization\cite{su}.
 
   The existence of localized $S=1/2$ state between two regions
with different polarization of dimerization can also be understood 
from the topological consideration of the $NL\sigma{M}$. In this
description, the spin chain is equivalent to a $\sigma$-model
with different topological angles $\theta_L=\pi(1-\gamma)$
and $\theta_R=\pi(1+\gamma)$ on two sides of the spin chain, 
where $\gamma>0$. The qualitative
properties of this model can be undertsood 
using a large-N expansion in the $CP_{N-1}$ representation
of the $NL\sigma{M}$, where spin excitations are represented
by charged-bosons in a one-dimensional universe of
scalar QED\cite{cole},
\begin{equation}
{\em L}\sim{1\over{g}}\left[|(\partial_{\mu}+iA_{\mu})Z|^2+\Delta_g^2
|Z|^2\right]+{1\over4e^2}F^2_{\mu\nu}-i{\theta(x)\over2\pi}
\epsilon^{\mu\nu}\partial_{\mu}A_{\nu},
\label{cpn}
\end{equation}
where $Z=\left[
\begin{array}{c}
Z_{\uparrow} \\
Z_{\downarrow}
\end{array}\right]$ is a two component spinor boson field, $F_{\mu\nu}=
\partial_{\mu}A_{\nu}-\partial_{\nu}A_{\mu}$. $e$ is the electric
charge carried by the bosons and the sign of the charge is in fact a
sub-lattice index\cite{ng94}. The $\theta$-term is given by 
$\theta(x)=\theta_{L(R)}$ on left(right) sides of spin chain and
is interpolating smoothly between the two regions. In this
description, the $\theta$-term gives rise to a background
electric field $E_b(x)=(\theta(x)/2\pi)e$, and the ground state
is obtained by minimizing the energy of the system in the
presence of $E_b(x)$\cite{cole}. The electric field energy of
the system can be reduced by
nucleating a boson of charge $-e$ out from vacuum and
localized at the boundary region where $\gamma$
changes sign. The total electric field is the sum of the
background electric field and electric field radiated from the 
charge boson. The resulting electric field remains unchanged
at the left-hand region, but is reduced to $E=((1+\gamma)/2-1)e=
-(1-\gamma)e/2$ on the right-hand region. The resulting change in energy
is of order $\Delta_g-e\gamma(L/2)$, where $\Delta_g$ is the energy
cost in nucleating the boson, and $L$ is the size of system. 
The energy of the system is always reduced for large enough $L$. 
The nucleated boson in this theory is precisely the 
localized $S=1/2$ state at the middle of spin chain. 

   The localized spin states between regions of spin
chain with different topological angle $\theta$ can also
be used to understand the local moments generated in the
$CuGe_{1-y}Si_yO_3$ compound. In this case, the spin chain is not
broken into open chains by $Si$. However, replacing $Ge$ by
$Si$ changes the coupling between spins close to the $Si$ ion,
or the elastic constant governing the lattice distortion, and may
lead to a local broken symmetry where one polarization of
dimerization is favoured over the other for the region close
to the $Si$ impurity. For finite concentration of $Si$ randomly
distributed in space, the favoured polarization is randomly
distributed, and the system can be described as a dimerized
spin chain with domain walls formed between 'mismatch' $Si$ 
impurities, or equivalently, a $NL\sigma{M}$
with topological angle $\theta(x)=\pi(1\pm\gamma)$
randomly distributed in space. Localized $S=1/2$ state will 
be formed at boundaries between regions with different $\theta$
as demonstrated above.

  The appearance of localized magnetic moments leads to
Curie behaviour in uniform susceptibility $\chi\sim(x,y)/2T$ 
at low temperature $T<\Delta_g$ and low doping, for both
$Cu_{1-x}Zn_xGeO_3$ and $CuGe_{1-y}Si_yO_3$ compounds.
Notice that the concentration of magnetic moment is on
average, half of the concentration of impurities in both cases.
At lower temperature and higher concentration of impurities, 
interaction between magnetic moments becomes dominant
and new magnetic states may develope in the systems.
In the following we shall derive the effective Hamiltonian
for the impurity-induced magnetic moments.
To derive the effective Hamiltonian, we divide
our system into two parts, the bulk, dimerized system and
the impurity-induced local moments, the bulk system is
described by the Hamiltonian with the local magnetic
moments removed, and the total Hamiltonian is
\[
H=H_o+J'\sum_{<i',j>}\vec{S}_{i'}.\vec{S}_{j},  \]
where $H_o$ is the Hamiltonian for the bulk system, $\vec{S}_{i'}$ are
the localized moments, $j$'s are the sites in the bulk system
next to $i'$, and $J'<J$ is some effective coupling between
the local moments and spins in the bulk
system. The effective interaction between the local
moments can be obtained by integrating out the bulk system. 
To second order in $J'$, we obtain an effective interaction between
local moments coming from exchange of bulk spinwaves,
\begin{mathletters}
\label{heff}
\begin{equation}
\label{hei1}
H_{eff}=\sum_{i',j'}J_{i'j'}\vec{S}_{i'}.\vec{S}_{j'},
\end{equation}
where $\vec{S}_{i'}$ and $\vec{S}_{j'}$ represent the localized 
magnetic moments, and\cite{si}
\begin{equation}
\label{hei2}
J_{i'j'}\sim-(J')^2\sum_{i,j}<\vec{S}_i.\vec{S}_j>,
\end{equation}
where $i$'s and $j$'s are the sites next to the local moments $i'$
and $j'$, and $<\vec{S}_i.\vec{S}_j>$ is the zero-frequency spin-spin
correlation function between sites $i$ and $j$ in the bulk system.
In particular, for large distance $r_{ij}=|\vec{r}_i-\vec{r}_j|$
and in the spin-Peierls state, we expect 
\begin{equation}
<\vec{S}_i.\vec{S}_j>\sim(-1)^{i+j}e^{-r_{ij}\over\xi},
\label{hei3}
\end{equation}
\end{mathletters}
where $\xi\sim\Delta_g^{-1}$ is the correlation length of the bulk spin
system, and $<S_iS_j>$ is positive when $i$ and $j$ are on the same
sublattice, and is negative otherwise, because of the underlying
antiferromagnetic correlation in the system. As a result,
$J_{i'j'}$ is negative between local moments on the same
sublattice, and is positive otherwise. Notice that intra-chain
coupling can also be included in the effective Hamiltonian by
including intra-chain coupling in the bulk Hamiltonian $H_o$. 
The qualitative result \ (\ref{heff}) is not modified except that
the correlation length $\xi$ is anisotropic with a much larger
value along chain and smaller value between chains.

   Now we may apply Eqs.\ (\ref{heff}) to the $Cu_{1-x}Zn_xGeO_3$
and $CuGe_{1-y}Si_yO_3$ systems. For the $Zn$-doped system,
inter-chain coupling between localized moments is extremely small
because the spin chain is broken, and the induced local moments
are located at different segments of the broken chain.
The ground state behaviour of the system is mainly determined
by intra-chain coupling, resulting in an effective
random (but unfrustrated) Heisenberg antiferromagnet model in
3D. The coupling $J_{ij}$ is random because the locations of the
local moments are random. However the system is unfrustrated 
because although the magnitude of the coupling
$J_{ij}$'s are random, the {\em sign} of the effective 
interaction is not: local moments on the same sublattice
always interact ferromagnetically whereas local moments
on different sublattice always interact antiferromagnetically.
The classical ground state of the impurity system is
always the Neel state which is unfrustrated.

   For the $Si$-doped system, the consideration is similar except
that inter-chain interaction between local moments exists and
dominate the physics at an intermediate temperature regime
when intra-chain coupling is not as important. Because of the
topological nature of the domain walls, 
we find that the local moments are always arranged in a 
staggered way where the local moments adjacent to each other
are always locate on opposite sub-lattice and interact
antiferromagnetically, i.e. the impurity-induced local
moments can be described by a model of 1D
antiferromagnetic spin chain with nearest neighbor
interaction $J_{ij}$ which is {\em always positive} but with random
magnitude. At lower temperature when intra-chain coupling
becomes important, the system is described by an effective
3D random Heisenberg antiferromagnet as in $Zn$-doped case.

  The theoretically suggested similarity and difference 
between the $Zn$-doped and $Si$-doped $CuGeO_3$ compounds
seem to be in qualitative agreement with what is observed
experimentally. At very low temperature when intra-chain coupling
becomes important, both systems are described by a model of
3D random Heisenberg antiferromagnet which, in the continuum
limit, can be described by a (3+1)d $NL\sigma{M}$ with random
spin-wave velocity $c(\vec{x})$ and random spin stiffness
$\rho_s(\vec{x})$,
\[
{\em S}={1\over2\hbar}\int{d}\tau\int{d}^3x\rho_s(\vec{x})\left[
|\nabla\vec{n}|^2+{1\over{c}(\vec{x})^2}\left({\partial\vec{n}\over
\partial\tau}\right)^2\right],   \]
and has an ordered
phase when the effective randomness is weak enough. This
long-range ordered antiferromagnetic phase seem to be observed
in both compounds\cite{oser,neu,pha}. A crucial difference
between the two compounds is that in the $Zn$-doped compound,
the AF phase is observed at a much larger value of doping,
when $x\sim0.03$, where the spin-gap is already unobservable,
\cite{ed1,oser}; whereas in the case of $Si$-doped compound, the
AF phase is observed at $y\sim0.007$, with co-existence of
AF phase and spin-Peierls phase observed\cite{neu}.
The much larger value of doping required in generating a AF
state in the $Zn$-doped compound is consistent with the absence
of inter-chain coupling between impurity-induced local moments
in $Zn$-doped compound. The disappearance of spin-Peierls
order in the AF phase also suggests that the AF phase may
be formed by ordering of {\em all} spins in the $Zn$-doped compound, 
but not ordering of impurity-induced local moments. Notice that 
co-existence of long-range antiferromagnetic and spin-Peierls order
in the $Ge$-doped compound is a natural prediction of our
theory if the AF state is formed by ordering of impurity-induced
local moments, since the local magnetic moments can exist only in
the presence of spin-Peierls order in our theory.

  It is also interesting to compare the low-energy properties of
$Si$-doped spin-Peierls and the $Zn$-doped two-leg ladder compounds.
In particular, interesting differences between the two compounds is
expected at the intermediate temperature regime, when the 
intra-chain coupling is not important. In this case, the
low energy physics of both compounds can be described by 
model of $S=1/2$ Heisenberg spin chain with random nearest
neighbor interaction, with one crucial difference: in the case of
$CuGe_{1-y}Si_yO_3$, the effective nearest neighbor interaction has 
random magnitude but is always positive, whereas in the $Sr(Cu_{1-x}Zn_x)_2
O_3$ compound, the effective nearest neighbor interaction 
has both random magnitude and sign\cite{si}. 
In the first case, the ground state of the system is
described by a random singlet phase\cite{dm,df}
which is characterized by formation of singlets between spins that
may be far apart from each other. In the second case, effective
spins with magnitude $>1/2$ are generated at low energy because of
ferromagnetic coupling\cite{wf} and drives the system towards a
fix point very different from the above\cite{wf}.
For example, it was found that the uniform magnetic susceptibility
$\chi\sim{T}^{-1}$ in the random sign and magnitude model
with different Curie
constant at low and intermediate temperatures\cite{si} 
whereas $\chi\sim{T}^{-1}|lnT|^{-2}$ in the random singlet
phase. The specific heat $C$ is also found to be very different
in the two cases, with $C\sim|lnT|^{-3}$ in the random singlet
phase and $C\sim{T}^{2\alpha}|lnT|$ in the second case with
$\alpha\sim0.22$\cite{wf}. Experimental tests of these differences
are suggested.

  The interesting low energy behaviour associated with impurity-doped
quasi- one dimensional spin-gap systems lead us to ask whether
similar interesting phenomena can be observed in other 
quasi-1D spin systems. Indeed, it has been suggested
that similar interesting physics can be observed in
half-integer spin chains with $S>1$\cite{ng94}, which
have bulk low energy physics identical
to $S=1/2$ spin chains\cite{hal,aff}. It was predicted
that local moments coupled to each other with
effective interaction $J'\sim1/L|lnL|$ will form at ends of
{\em open} chains of length $L$.\cite{ng94,qsj}. Following similar
analysis as in Refs\cite{si,ng94}, it can be shown that 
analogous low energy behaviour can be observed also in the
three-leg Heisenberg ladder compound, upon substitution of 
$Cu$ by $Zn$. It is convenient to view the three-leg ladder 
compound as three coupled antiferromagnetic spin chains.
We find that replacement of $Cu$ at the central chain
of the three-leg ladder compound by $Zn$ results at different
physics from when the $Cu$ at one of the side-chains is
replaced\cite{ng96}. First we consider replacing one single $Cu$
atom by $Zn$. In the case when the $Cu$ is located at one
of the side-chains, we find that the $Zn$ impurity
effectively 'breaks' the ladder and the low energy physics
of the system is the same as two uncoupled $S=1/2$ spin chains; whereas
in the case when the $Cu$ atom is located at the central chain,
we find that the $Zn$ atom introduces effectively a $S=1$ 
impurity into the spin-ladder and the low energy physics is
that of two $S=1/2$ spin chains 
coupled antiferromagneticaly to the $S=1$
impurity spin from both side. For finite concentration of
$Zn$ impurities, the low energy physics of the system can be 
described by $S=1/2$ spin chains coupled to randomly
located $S=1$ impurities (assuming that at least
part of the $Cu$ atoms at central chain are replaced by $Zn$) and
the low energy behaviour of the system is determined by 
competition between Kondo screening of the impurity spins
by the efffective $S=1/2$ spin chain\cite{ea} and antiferromagnetic 
ordering of impurity-induced $S=1$ local moments, which may lead
to very different low temperature behaviour compared with the
spin-Peierls or two-leg ladder systems\cite{ng96}.

   I acknowledge helpful discussions with H. Fukuyama, A. Azuma, 
N. Nagaosa, A. Furusaki and M. Sigrist. I also thank the 
Yukawa Institute, Kyoto, where part of this work was done.


\begin{references}
 \bibitem{edimer} M. Hase, I. Terasaki and K. Uchinokura, \prl
 {\bf 70}, 3651(1993).
 \bibitem{el2} Z. Hiroi, M. Azuma, M. Takano and Y. Bando, J.
 Solid State Chem. {\bf 95}, 230(1991).
 \bibitem{rice} T.M. Rice, S. Gopalan and M. Sigrist, Europhys. Lett.
 {\bf 23}, 445(1993).
 \bibitem{ed1} M. Hase {\em et.al.}, \prl {\bf 71}, 1450(1995).
 \bibitem{ed2} J.P. Renard {\em et.al.}, Europhys. Lett. {\bf 30},
 475(1995).
 \bibitem{oser} S.B. Oseroff {\em et.al.}, \prl {\bf 74}, 1450(1995).
 \bibitem{neu} L.P. Regnault, J.P. Renard, G. Dhalenne and A.
  Revcolevschi, Europhys. Lett. {\bf 32}, 579(1995).
 \bibitem{pha} M. Poirier {\em et.al.}, \prb {\bf 52}, R6971(1995).
 \bibitem{az1} M. Azuma  {\em et.al.}, \prl {\bf 73}, 3463(1994).
 \bibitem{az2} M. Azuma {\em et.al.}, preprint.
 \bibitem{fu} H. Fukuyama, T. Tanimoto and M. Saito, J. Phys.
   Soc. Jpn. {\bf 65}, 1183(1996).
 \bibitem{na} H. Fukuyama, N. Nagaosa, M. Saito and T. Tanimoto,
  J. Phys. Soc. Jpn. {\bf 65}, 8(1996).
 \bibitem{si} M. Sigrist and A. Furusaki, J. Phys. Soc. Jpn.
 {\bf 65}, 5901(1996).
 \bibitem{hal} F.D.M. Haldane, Phys. Lett. {\bf 93A}, 464(1983).
 \bibitem{aff} see I. Affleck, in {\em Field Theory Methods and
 Quantum Critical Phenomena}, edited by E. Brezin and J. Zinn-Justin
 (North-Holland, Amsterdam, 1990).
 \bibitem{cf} M.C. Cross and D.S. Fisher, \prb {\bf 19}, 402(1979).
 \bibitem{ng94} T.K. Ng, \prb {\bf 52}, 555(1994).
 \bibitem{su} W.P. Su, J.R. Schrieffer and A.J. Heeger, \prl
  {\bf 42}, 1698(1979).
 \bibitem{cole} S. Coleman, Ann. Phys. (N.Y.) {\bf 101}, 239(1976).
 \bibitem{dm} C. Dasgupta and S.K. Ma, \prb {\bf 22}, 1305(1980).
 \bibitem{df} D.S. Fisher, \prb {\bf 50}, 3799(1994).
 \bibitem{wf} E. Westerberg, A. Furusaki, M. Sigrist and P.A. Lee,
  \prl {\bf 75}, 4302(1995).
 \bibitem{qsj} S.J. Qin, T.K. Ng and Z.B. Su, \prb {\bf 52}, 12844
 (1995).
 \bibitem{ng96} T.K. Ng, unpublished.
 \bibitem{ea} S. Eggert and I. Affleck, \prb {\bf 46}, 10866(1992).
\end{references}
\end{document}